\documentclass[square]{ws-procs11x85}
\usepackage{balance} %necessary when used with \twocolumn

\def\Journal#1#2#3#4{{#1} {\bf #2}, #3 (#4)}

\newcommand{\met}{\hbox{E\kern-0.5em\lower-0.1ex\hbox{/}}_T}

\begin{document}

\twocolumn[
\title{Gamma-ray emission of relativistic jets as a supercritical process}

\author{B. E. Stern $^{1,2,3}$ and J. Poutanen$^{3}$}

\address{$^{1}$ Institute for Nuclear Research, Russian Academy of Sciences,
Prospekt 60-letiya Oktyabrya 7a, Moscow 117312, Russia} 

\address{$^{2}$ Astro Space Center, Lebedev Physical Institute,
Profsoyuznaya 84/32,  Moscow 117997, Russia \\
E-mail: stern@bes.asc.rssi.ru}  

\address{$^{3}$ Astronomy Division, Department of Physical Sciences, P.O.Box 3000, 90014 University of Oulu, Finland \\ E-mail:  juri.poutanen@oulu.fi}

%%%%%%%%%%%%%%%%%%%%%%%%%%%%%%%%%%%%%%%%%%%%%%%%%%%%%%%%%%%%%%%%%%%%%%%%%
% You may repeat \author \address as often as necessary                 %
%%%%%%%%%%%%%%%%%%%%%%%%%%%%%%%%%%%%%%%%%%%%%%%%%%%%%%%%%%%%%%%%%%%%%%%%%

\begin{abstract}
Supercriticality of the same kind as that in a nuclear pile
can take place in high-energy astrophysical objects producing a number of
impressive effects. For example, it could cause an explosive release of the
energy of a cloud of ultrarelativistic protons into radiation. More certainly,
supercriticality should be responsible for energy dissipation of very
energetic relativistic fluids such as ultrarelativistic shocks in gamma-ray bursts
and jets in active galactic nuclei (AGNs).
In this case, the photon breeding process operates. It is a kind of the converter 
mechanism  with the high-energy  photons and $e^+e^-$ pairs converting into each other 
via pair production and inverse Compton scattering.
Under certain conditions, which should be satisfied in powerful AGNs,
the photon breeding mechanism becomes supercritical: the high-energy photons
breed exponentially until their feedback on the fluid changes its velocity
pattern. Then the system comes to a self-adjusting near-critical steady
state. Monte-Carlo simulations with the detailed treatment of particle
propagation and interactions demonstrate that a jet with the Lorentz
factor $\Gamma\approx 20$ can radiate away up to a half of its total energy and for $\Gamma=40$
the radiation efficiency can be up to 80 per cent. Outer layers of the jet decelerate down to a
moderate Lorentz factor 2--4, while the spine of the jet has the final
Lorentz factor in the range 10--20 independently on the initial  $\Gamma$. Such sharp
deceleration under the impact of radiation must cause a number of
interesting phenomena such as formation of internal shocks and an early generation
of turbulence.
\end{abstract}
\vskip12pt  % insert '\vskip12pt' while using '\twocolumn' command
%\vskip28pt % if there is no keywords
]

\bodymatter

\section{Supercritical processes in high-energy astrophysics}

The term ``supercriticality'' has different meanings. Here we use
the same meaning as in nuclear physics: the condition of exponential runaway
behaviour of the system (nuclear explosion or nuclear pile). 
So far  two kinds of supercritical phenomena   were proposed in 
high-energy astrophysics: an explosive energy release of a cloud of 
ultrarelativistic protons into photons (``proton bomb'') and dissipation of 
the bulk energy of relativistic fluids into radiation.
 
Let us consider a magnetized cloud of relativistic protons with typical Lorentz factor $\gamma_p$.
These protons do not radiate much, 
because the synchrotron radiation and the inverse Compton scattering rates for protons are 
suppressed by factor $(m_e/m_p)^2$ with respect to electrons. Suppose 
$N$ seed photons with energy $\varepsilon_t \sim 200{\rm MeV}/\gamma_p$
interact with protons producing pions decaying into high-energy 
electrons or photons. They, in turn, produce a wide spectrum of softer photons, 
with the number $\xi N$ of them on average interacting with the high-energy protons again.  
If  $\xi > 1$  then the process has a supercritical character: the number of photons rises 
exponentially until the protons cool down, so that the supercriticality condition 
breaks down and the energy release terminates. If the energy supply to protons 
continues then the system reaches supercriticality and explodes again.
Such scenario was proposed and studied with numerical simulations \cite{SS91} and analytically \cite{KM92}. 

The supercriticality condition depends on the radiation spectrum emitted by the particles 
produced in photo-meson interaction. The hardest possible spectrum is the
fast-cooling power-law $dN/d\varepsilon \propto \varepsilon^{-3/2}$. The softest 
spectrum is that of the saturated pair cascade  $dN/d\varepsilon \propto \varepsilon^{-2}$ \cite{sve87},
which takes place in a dense soft photon field.
 
The cross-section of photo-meson production multiplied by the fraction of energy transfered to pions, $x$, 
is (see \cite{ber01}) $\sigma (p\gamma) x \sim 0.7\times 10^{-28} {\rm cm}^2$ in the photon energy range 
$\sim 250$ MeV--2 GeV in the proton rest frame. The number of produced photons which can interact 
again is $N \sim 1/\gamma_p$ in the first case and $N \sim 1/\gamma_p^2$ in the second case. Then condition $\xi > 1$ translates to $\rho_E > 10^{25}$ erg cm$^{-2}$
in the fast cooling case and  $\rho_E > 10^{25}/\gamma_p$ erg cm$^{-2}$ in the 
cascade case, where $\rho_E$ is the column energy density of protons across the cloud.

Where one can find such a cloud of ultrarelativistic protons? 
%The papers cited above  have considered the phenomenon in general. 
%It was just assumed that it could be relevant for  the AGN nonthermal emission. 
A possible site for this phenomenon, gamma-ray bursts (GRBs), was proposed by  \cite{KGM02}.
Indeed, all scenarios of GRBs involve ultrarelativistic shocks, where the space between 
the shock front and the contact discontinuity is filled with ultrarelativistic protons. 

 Another possible class of supercritical phenomena can take place in 
relativistic flows. A first example  is the  supercritical pair loading  
in AGN jets near  the accretion disc's hot corona which emits hard X-rays \cite{Lev96}.
If the jet Lorentz factor at such distance is 
$\Gamma \sim 10$, then cool pairs moving with the jet can up-scatter   
X-rays to MeV energies and these photons in their turn, interacting with the same X-rays,
can produce new pairs.  

 A more powerful and universal way of dissipation of relativistic fluids
into radiation, is associated with the converter mechanism
suggested in general form by Derishev et al. \cite{Der2003}  and  in a 
more specific form of a runaway (supercritical) electromagnetic cascade in ultrarelativistic 
shocks by Stern \cite{St2003}. We later performed a numerical study of the 
converter mechanism, in its electromagnetic version, operating in relativistic jets 
of blazars  \cite{SP06,SP08} and found that it has a supercritical character: the high-energy
photons and relativistic $e^+e^-$ pairs breed exponentially.

\section{Converter mechanism and photon breeding} 

 The converter mechanism is an alternative to the Fermi acceleration for
relativistic flows in a dense radiation field. In the Fermi mechanism a charged 
particle gains energy being scattered many times between media moving 
with respect to each other. In the converter mechanism, a quantum of energy 
moves between media in a form of a neutral particle and scatters taking a form of a 
charged particle. The conversion of a particle is due to interaction with a soft 
background radiation. In the case of an ultrarelativistic flow, the converter mechanism 
can be more efficient than Fermi acceleration because a neutral particle can more easily 
cross the boundary between the flow and the external environment. 

There can be two types of charge exchange cycles: (1) 
proton--neutron cycle with proton conversion into neutron via photo-meson production and subsequent neutron decay, (2) photon--electron/positron cycle with conversion due to photon-photon pair 
production and inverse Compton scattering.

 The important feature of the second cycle is the absence of any limitation of the 
number of participating particle such as the baryon number conservation in the first case.
High-energy photons and pairs can breed exponentially     
at certain conditions being fed by the bulk energy of the flow \cite{St2003,SP06,SP08}.  
  
A very simplified description of the photon breeding cycle for the case 
of a relativistic jet can be represented by five steps.

 (i) An external high-energy photon of energy $\varepsilon$ enters the jet and interacts with a soft background photon 
producing an electron-positron pair.

(ii) The time-averaged Lorentz factor of the produced pair (as measured in the external frame) 
gyrating in the  magnetic field of the jet, becomes $\gamma\sim \Gamma^2 \varepsilon$.

(iii) The electrons and positrons in the jet Comptonize soft photons (internal synchrotron or external)  
up to high energies.

(iv) Some of these photons leave the jet and produce pairs in the external environment. 
 
(v) Pairs gyrate in the magnetic field and Comptonize soft photons more or less
isotropically.  Some of these Comptonized high-energy photons enter the jet again.

In this cycle, the energy gain, $\sim \Gamma^2$, is provided by the isotropization of the charged particles in the jet frame and is taken from the bulk energy of the flow. 
Other steps in the cycle are energy sinks.  
The whole process proceeds in a runaway regime, with 
the total energy in photons and relativistic particles increasing exponentially, if 
the amplification coefficient (energy gain in one cycle) is larger than unity. 

This new mechanism is actually much simpler in the description than 
Fermi acceleration, which depends on detailed geometry of the magnetic field and the 
poorly understood supply of seed nonthermal particles. 
Once the velocity pattern for the fluid, the magnetic field and the external 
radiation field are specified, the fate of each high-energy photon and its descendants 
can be reproduced (in statistical sense) from first principles, 
because the interaction cross-section are known with high accuracy. 
The question whether a photon produces a runaway avalanche or not, can be answered exactly. 

\section{Dissipation of jet bulk energy into radiation}

Stern \& Poutanen \cite{SP06, SP08} performed a detailed 
numerical simulation of the photon breeding in AGN jets. The simulation of the 
particle emission propagation and interactions was exact and based on   first 
principles, while the fluid dynamics was treated with a simplified 
2D ballistic approximation. Here we present some results 
obtained under assumption that the main emission site is the broad emission line region  \cite{SP08}. The 
soft photon background which is necessary for the converter mechanism was composed 
from direct radiation of accretion disc and an isotropic component, which includes reprocessed/scattered
radiation of the disc and surrounding dust.

\begin{figure}[t]
\center
\centerline{\psfig{figure=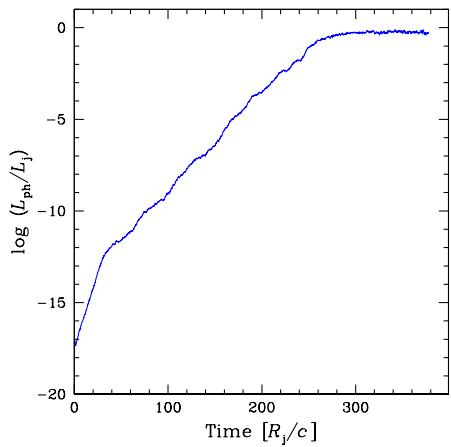,width=6truecm}}
\caption{Fraction of the jet power converted into photons versus time. 
The initial density of the high-energy, seed photons corresponds to the 
extragalactic gamma-ray background. Parameters: distance from the central engine $R = 2 \times 10^{17}$ cm, 
jet radius $R_j = 10^{16}$ cm, jet and disc luminosities 
$L_j = L_d = 10^{44}$ erg s$^{-1}$, and Poynting flux of $L_B\approx10^{43}$ erg s$^{-1}$. From \cite{SP08}.}
\label{fig:lows}
\end{figure}

 To initiate the photon breeding one needs a number of seed high-energy photons. Their 
origin is not important as their number can be arbitrarily small. Fig. 1 represents 
a simulation starting from the seed gamma-rays corresponding to the extragalactic gamma-ray
background. The energy release increased by 20 orders of magnitude during 
$250 R_j/c$ ($\sim 3$ years for the jet radius $R_j = 10^{16}$ cm) and came 
to the steady state at the level $\sim 0.5$ of the total jet bulk energy.   
The rapid rise at small times corresponds to the exponential growth of the photon avalanche
as it moves downstream with the flow. Then as the avalanche reaches the end of the ``simulation volume'' 
(of length $20 R_j$) it cannot grow further, and the growth is further supported by the up-streaming photons,  
which provide a spatial feedback loop.

\begin{figure}[t]
\center
\centerline{\psfig{figure= 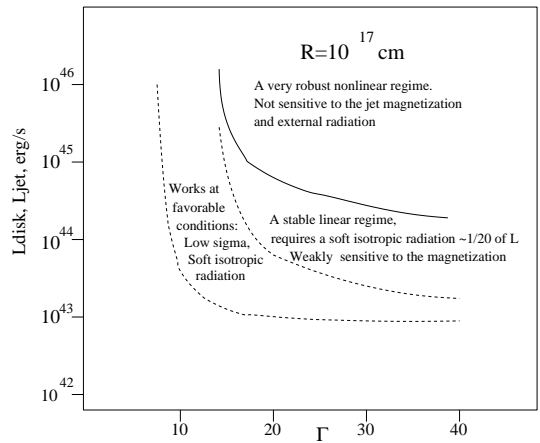,width=7.0truecm}}
\caption{Working areas of the photon breeding mechanism in luminosity--Lorentz factor plane. Solid curve refers to the jet power, dashed curves -- to the disc luminosity.}\label{fig:reg}
\end{figure}

We have performed several tens simulation runs \cite{SP08} and explored the parameter space, where 
the photon breeding can work. A schematic representation of the results is given in Fig. \ref{fig:reg}.
The minimal jet Lorentz factor  for the supercritical breeding was $\Gamma = 8$.
At a moderate Lorentz factor the mechanism works only at favourable conditions.  
One of this conditions is weak magnetic field. 
The magnetic energy flux of the jet should be much smaller than the accretion disc luminosity. 
Otherwise the synchrotron losses by pairs at step (iii) of the breeding cycle dominate 
over Compton losses. The synchrotron photons are too soft to produce pairs and do not 
participate in the breeding, reducing its efficiency. 
If the jet has low magnetization ($\sigma \ll 1$) 
or the jet power  is smaller than the disc luminosity, the breeding is favoured. 

The second favourable condition is the presence of a soft isotropic radiation field. The 
broad line photons and scattered photons from the disc are not sufficiently soft, because pairs 
in the jet at step (iii) interact with them in a deep Klein-Nishina regime. 
Therefore again synchrotron radiation reduces the breeding efficiency. 
A softer radiation can be supplied by the surrounding dust or by the jet itself. 

The third condition is a high accretion disc luminosity, $L_d > 10^{43}/R_{17}$ erg s$^{-1}$
(where $R_{17} = R/10^{17}$cm is the distance from the disc to the site of the photon breeding), 
which is necessary to provide sufficient photon-photon opacity across the jet.

At a higher Lorentz factor the requirement of a weak magnetic field is relaxed. 
If $\Gamma > 20$, then a magnetically dominated jet with $L_j\sim L_d$ 
can radiate efficiently. The above conditions are relevant 
for the linear case, when the jet power is moderate and the soft synchrotron 
radiation of the jet is optically thin for gamma-rays. The non-linear effects 
appear at the jet power $L_j \sim 3\times 10^{44}$ erg s$^{-1}$ (for $R \sim 10^{17}$cm). 
At $L_j  > 10^{45}$ erg s$^{-1}$ the radiative steady-state becomes 
self-supporting even without external radiation.

\begin{figure}[t]
\center
\centerline{\psfig{figure= 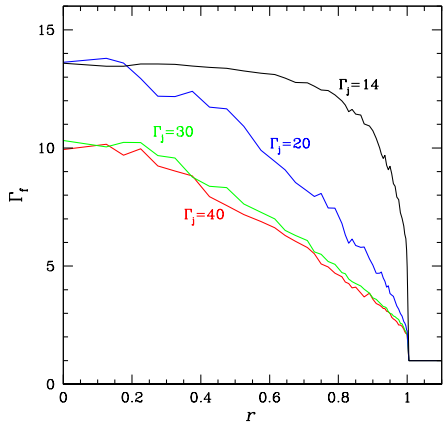,width=6truecm}}
\caption{Terminal Lorentz factor $\Gamma_{\rm j}$ at the outlet of the cylindric ``simulation volume'' with length $20 R_j$ 
versus distance from the jet axis for different initial Lorentz factor of the jet.
}\label{fig:lor}
\end{figure}

The radiative efficiency of the jet increases with $\Gamma$ 
and reaches  0.14, 0.56, 0.77, 0.82 for  $\Gamma = 14, 20, 30, 40$, respectively \cite{SP08}. 
A high efficiency implies that the jet undergoes strong deceleration.
The distribution of terminal $\Gamma$ across the jet  is shown in Fig. \ref{fig:lor}. 
One can see that the deceleration is very inhomogeneous: the final 
Lorentz factor at the jet boundary is 2--4, while at the center its value is 10--14.

\begin{figure}[t]
%\center
\centerline{\psfig{figure= 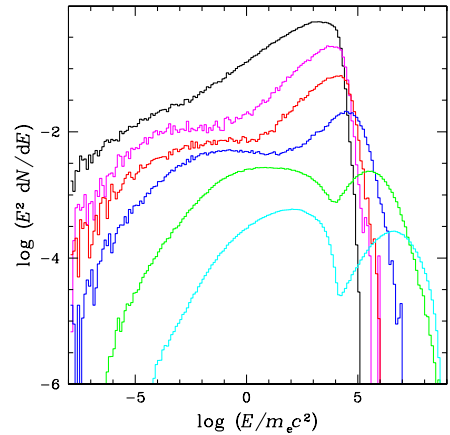,width=6truecm}}
\caption{Simulated spectra for the case $L_j = L_d$, $L_B = 0.2 L_j$ and 
$L_d$ varying from $5\times10^{43}$ to $10^{46}$ erg s$^{-1}$.
}\label{fig:spec}
\end{figure}

The resulting spectral energy distributions for a sequences of runs with 
varying luminosity are shown in Fig. \ref{fig:spec}.
We see a tendency similar to the observed ``blazar sequence'': spectra
for higher luminosity are shifted to lower energy. 
The observed blazar spectra demonstrate two distinct 
components traditionally interpreted as synchrotron and inverse Compton 
emission peaks of the same electrons.  In our simulations the synchrotron and 
Compton components are broad and overlap.  The possible reason for this 
difference is that the photon breeding mechanism works together with 
other mechanisms like diffusive or internal shock acceleration.
Moreover, the photon breeding produces a strong impact on the jet, after the 
rapid inhomogeneous deceleration the jet should be highly perturbed. In our 
interpretation the soft blazar component is emitted further downstream 
due to a secondary process, e.g., diffusive reheating of cooled pairs produced 
by the photon breeding mechanism.

\section*{Acknowledgments}
This work is supported by RFBR grant 07-02-00629-a, 
Ehrnrooth and  V\"ais\"al\"a Foundations, and  
the Academy of Finland grants 110792 and 112982.

\balance

\end{document}